\documentclass{conm-p-l}

\newtheorem{theorem}{Theorem}[section]

\newtheorem{corollary}[theorem]{Corollary}

\numberwithin{equation}{section}
\numberwithin{figure}{section}

\begin{document}

\title[Characteristic Cauchy problem]{An existence theorem for the Cauchy
problem on a characteristic cone for the Einstein equations}

\author{Yvonne Choquet-Bruhat}
 \address{Acad\'emie des Sciences, Paris}

\author{Piotr T. Chru\'{s}ciel}
 \address{F\'{e}d\'{e}ration Denis Poisson, LMPT, Tours; Hertford
College and  Oxford Centre for Nonlinear PDE, University of
Oxford} \curraddr{Gravitationsphysik, Universit\"at Wien}
 \thanks{Supported in
part  by the EC project KRAGEOMP-MTKD-CT-2006-042360, by the
Polish Ministry of Science and Higher Education grant Nr N N201
372736, and by the EPSRC Science and Innovation award to the
Oxford Centre for Nonlinear PDE (EP/E035027/1).}

\author{Jos\'e M.
Mart\'in-Garc\'ia}
 \address{Laboratoire Univers et Th\'eories, Meudon, and\\
Institut d'Astrophysique de Paris}
 \thanks{Supported by the French ANR grant
BLAN07-1\_201699 entitled ``LISA Science'', and also in part by
the Spanish MICINN project FIS2009-11893.}

\begin{abstract}
We prove an existence theorem for the Cauchy problem on a
characteristic cone for the vacuum Einstein equations.
\end{abstract}

\maketitle


\section{Introduction}

In recent work~\cite{CCM2} (see also \cite{CCMWascom,CCG}) we
have analysed some aspects of the Cauchy problem for the
Einstein equations with data on a characteristic cone in all
dimensions $n+1\geq 3$,
see~\cite{YvonneCIVP,DossaAHP,F1,RendallCIVP2} and references
therein for previous work on the subject. In this note we apply
the results derived in~\cite{CCM2} to present an existence
theorem for this problem, with initial data which approach
rapidly the flat metric near the tip of the light cone, see
Theorem~\ref{T23VII.1} below.

The reader's attention is drawn to~\cite{FriedrichCMP86},
where sets of unconstrained  data on a light-cone centered at
past timelike infinity are given in dimension $3+1$.

\section{Free initial data with asymptotic expansions at the vertex}
 \label{sSolAFvert}

It is well known that by using normal coordinates centred at
$O$ the characteristic cone $C_O$ of a given Lorentzian metric
can be written, at least in a neighbourhood of $O$, as a cone
in Minkowski spacetime whose generators represent the null
rays. It is therefore no geometric restriction to assume that
the characteristic cone of the spacetime we are looking for is
represented in some  coordinates $y:=(y^{\alpha})\equiv
(y^{0}$, $y^{i}$, $i=1,\ldots,n)$ of $\mathbf{R}^{n+1}$ by the
equation of a Minkowskian cone with vertex $O$,
\begin{equation}
r-y^{0}=0,\quad  r:=\big\{\sum(y^{i})^{2}\big\}^{\frac{1}{2}}.
\end{equation}
The parameter $r$ is an affine parameter when normal
coordinates are used, and it is also going to be an affine
parameter in the solutions that we are going to construct.

Coordinates $y^\mu$ as above, which moreover satisfy the
wave-equation, $\Box_g y^\mu =0$, will be called
\emph{normal-wave} coordinates. Given a smooth metric, such
coordinates can always be constructed
(see~\cite{Friedlander} or~\cite{DossaAHP}) by solving
the wave equation in the domain of dependence of $C_O$, with
initial data the normal coordinates on $C_O$. Given a
smooth metric, one obtains a coordinate system near the vertex,
which suffices for our purposes. The coordinates $y$ will be
normal-wave coordinates for the solution which we aim to
construct.

Given the coordinates $y^{\alpha}$ we can define coordinates
$x^{\alpha}$  on $\mathbf{R}^{n+1}$ by setting
\begin{equation}
 \label{1IX.1}
x^{0}=r-y^{0}, \quad
x^{1}=r
 \;,
\end{equation}
with $x^{A}$ local coordinates on $S^{n-1}$.
The  null geodesics issued from $O$ have equation $x^{0}=0$,
{}$x^{A}=$constant, so that $\frac{\partial}{\partial x^{1}}$
is tangent to those geodesics. On $C_{O}$ (but not outside of
it in general) the spacetime metric $g$ that we attempt
to construct takes the form (we put an overbar to denote
restriction to $C_O$ of spacetime quantities)
\begin{equation}
 \label{null2}
\overline{g}:=g|_{x^{0}=0}\equiv \overline{g}_{00}(dx^{0})^{2}
+2\nu_{0}dx^{0}dx^{1}+2\nu_{A}dx^{0}dx^{A}+\tilde{g}
 \;,
\end{equation}
where
\begin{equation}
 \label{null4}
\nu_{0}:=\overline{g}_{01} , \quad
\nu_{A}:=\overline{g}_{0A} , \quad
\tilde{g}:=\overline{g}_{AB}dx^{A}dx^{B},\ \ A,B=2,\ldots,n,
\end{equation}
are respectively an $x^{1}$-dependent scalar, one-form, and
Riemannian metric on $S^{n-1}$. The symbol $r$ will be used
interchangeably with $x^1$.

To avoid ambiguities, we will write $g_{\mu\nu}$ for the
components of a metric tensor in the $x^\alpha$ coordinates,
and  $\underline{g_{\mu\nu}}$ for the components in the
coordinate system $y^\mu$. This convention will be used
regardless of whether $g$ is defined on the space-time, or only
on the light-cone.

The analysis in~\cite{CCM2} uses a wave-map gauge, with
Minkowski target $\hat g$, with the light-cone of $\hat g$
being the image by the wave-map $f$ of the light-cone of the
metric $g$ that one seeks to construct. Quite generally, a
metric $g$ on a manifold $V$ will be said to be \emph{in
$\hat{g}$-wave-map gauge} if the identity map $V\rightarrow V$
is a harmonic diffeomorphism from the spacetime $(V,g)$ onto
the pseudo-Riemannian manifold ($V,\hat{g})$.
Recall that a mapping $f:(V,g)\rightarrow (f(V),\hat{g})$ is a
harmonic map if it satisfies the equation, in abstract index
notation,
\begin{equation}
 \label{WaveMap}
\hat{\square}f^{\alpha} :=
  g^{\lambda \mu} (
    \partial_{\lambda\mu}^{2}f^{\alpha}
   -\Gamma_{\lambda\mu}^{\sigma} \partial_{\sigma}f^{\alpha}
   +\partial_{\lambda} f^{\sigma} \partial_{\mu} f^{\rho}
      \hat{\Gamma}_{\sigma\rho}^{\alpha}
  )=0.
\end{equation}
In a subset in which $f$ is the identity map defined by
$f^{\alpha}(y^\mu)=y^{\alpha}$, the above equation reduces to
$H=0$, where the  \emph{wave-gauge vector} $H$ is given in
arbitrary coordinates by the formula
\begin{equation}
 \label{WaveGauge0}
H^{\lambda} := g^{\alpha\beta} \Gamma_{\alpha\beta}^{\lambda}-W^{\lambda}
 \;,
\ \text{with}\
W^{\lambda} :=g^{\alpha\beta} \hat{\Gamma}_{\alpha\beta}^{\lambda}
 \;,
\end{equation}
where $\hat{\Gamma}^{\lambda}_{\alpha\beta}$ are the
Christoffel symbols of the \emph{target} metric $\hat{g}$.
See~\cite{YvonneBook} for a more complete discussion.

There are various ways of choosing free initial data for
the Cauchy problem for the vacuum Einstein equations on the
light-cone $C_O$. In this work we choose as initial data   a
one-parameter family, parameterized by $r$, of conformal
classes of metrics $[\gamma(r)]$ on $S^{n-1}$, thus $\tilde
g(r)$ is assumed to be conformal to $\gamma(r)$, and where
$r$ will be an affine parameter in the resulting vacuum
space-time. The initial data needed for the evolution
equations are the values of the metric tensor on the
light-cone, which will be obtained from $\gamma(r)$ by solving
a set of wave-map-gauge constraint equations derived
in~\cite{CCM2}, namely Equations (\ref{3V.1}), (\ref{5VI.13}),
(\ref{xiAbis})-(\ref{CAfinal2}) and (\ref{19VIII.1}) below. The
main issue is then to understand the behaviour of the fields
near the vertex of the light-cone, making sure that the
$y^\alpha$-coordinates components of the metric $\overline g$,
obtained by solving the wave-map-gauge constraints, can
be written as restrictions to the light-cone of sufficiently
smooth functions on space-time, so that the PDE existence
theorem of~\cite{DossaAHP}  can be invoked to obtain the vacuum
space-time metric.

It is convenient to start with some notation. For $\ell \in
\mathbf{N}$ and $\lambda \in\mathbf{R}$ we shall say that a
tensor field $\varphi(r,x^2,\ldots,x^n)$,  of valence $N$,
defined for $0<r\le r_0\le 1$, is $O_\ell(r^\lambda)$ if there
exists a constant $C$ such that
\begin{equation*}
\mbox{for all}\ j_1 +  \ldots+j_{n}\le \ell\ \mbox{we have} \
 |\partial_r^{j_1} \partial^{j_2}_{x^2}\ldots \partial^{j_{n }}_{x^{n }}
 \varphi_{i_1\ldots i_N}| \le C r^{\lambda-j_1}
 \;,
\end{equation*}
where $\varphi_{i_1\ldots i_N}$ are coordinate components of
$\varphi$, in a coordinate system which will be made clear as
needed.

Let $s_{AB}$ denote the canonical unit round metric on the
sphere $S^{n-1}$. We have:

\begin{theorem}
 \label{T23VII.1}
Let $m,n,\ell, N \in \mathbf{N}$, $n\ge 2$,
$m>n/2+3$, $\ell\ge 2m+2$, $N  \ge 2m+1$, $\lambda \in
[N,N+1]$. Suppose that
there exist smooth tensor fields $\stackrel{(i)}s_{\!AB}$ on
$S^{n-1}$, $i=4,\ldots  N$, so that, in local charts on
$S^{n-1}$, the coordinate components $\gamma_{AB}$ satisfy
\begin{equation}
 \label{30VI.5x}
 \gamma_{AB} - r^2 s_{AB} =
     \sum_{i=4}^N r^{i} \stackrel{(i)}{s}_{\!\!AB}+O_\ell(r^\lambda)
 \;.
\end{equation}
Then:
\begin{enumerate}
\item There exist functions
    $\underline{\stackrel{(i)}g_{\!\mu\nu}} \in
    C^\infty(S^{n-1})$ such that
\begin{equation*}
 \underline{\overline{g}_{\mu \nu}} - \underline{\eta _{\mu \nu}}
 = \sum_{i=2}^{N-2} r^i\underline{\stackrel{(i)}g_{\!\mu\nu}}
   + O_{\ell-3}(r^{\lambda-2})
 \;.
\end{equation*}
If there exists $4\le N_0\le N$ so that
$\stackrel{(i)}{s}_{\!\!AB}=0$ for $i=4,\ldots,N_0$, then
$\underline{\stackrel{(i)}g_{\!\mu\nu}}=0$ for $i=2,\ldots,N_0-2$.
\item If moreover
\begin{eqnarray}
 \label{23VII.5}
 &&
 \text{$\sum_{i=2}^{N-2} r^i\underline{\stackrel{(i)}g_{\!\mu\nu}}$
       are restrictions to the light-cone $C_O$}
\\
 \nonumber
 &&
 \text{
 of a polynomial in the variables $y^\mu$,}
\end{eqnarray}
then there exists a  $C^{m-(n-1)/2}$ Lorentzian metric
defined in a neighbourhood of the vertex of $C_O$, with
$\tilde g(r)\in [\gamma(r)]$, which is vacuum to the future
of $O$.
\end{enumerate}
\end{theorem}

Roughly speaking, the index $m$ is the final Sobolev
differentiability of the solution. The ranges of indices above
are only needed for the second part of the theorem, and arise
from the fact that Dossa's existence theorem~\cite{DossaAHP}
requires initial data which are of $C^{2m-1}$ differentiability
class in coordinates regular near the vertex. There is a  loss
of three derivatives when going from the free data
$\gamma_{AB}$ to the full initial data $\overline{g_{\mu\nu}}$,
which brings the threshold up to $2m+2$. Finally, the existence
argument invokes the Bianchi identity, which in its
natural version requires a $C^3$ metric; together with
the Sobolev embedding, this leads to the restriction $m>n/2+3$.

Straightforward Taylor expansions at $O$  show  that a smooth
metric on a space-time $(\mathcal M,g)$ will lead to the form
(\ref{30VI.5x}) of $[\gamma(r)]$, with $m$, $\ell$  and $N$
which can be chosen at will, and with $\lambda = N+1$. So
(\ref{30VI.5x}) is necessary in this sense.

When all the $\stackrel{(i)}{s}_{\!\!AB}$'s, $i=4,\ldots N$,
vanish, then so do the
$\underline{\stackrel{(i)}g_{\!\mu\nu}}$'s. It follows  that

\begin{corollary}
If $\gamma_{AB}$ approaches $r^2 s_{AB}$ as $r^{2m+1}$ or
faster, where $m$ is the smallest integer less than or equal to
$n/2+3$, then an associated solution of the vacuum Einstein
equations exists.
\end{corollary}

In dimension three the required rate is $\lambda \ge 11$.

In view of our theorem above, to obtain a complete
solution of the problem at hand it remains to provide an
exhaustive description of those $[\gamma(r)]$'s that lead to
(\ref{23VII.5}). We conjecture that (\ref{23VII.5}) will hold
for all $\gamma(r)$'s arising from the restriction of a smooth
metric to a light-cone, where $r$ is an affine parameter. A
similar problem arising on the null cone at past infinity in
dimensions $3+1$ has been solved by Friedrich
in~\cite{FriedrichCMP86}, but no details have been presented.

\bigskip

\noindent{\sc Proof of Theorem~\ref{T23VII.1}:} We need to
analyze the behaviour of the solutions of the
wave-map-gauge constraint equations near the vertex of
the cone. We start by noting that, for some smooth functions
$\stackrel{(i)}\psi$ on $S^{n-1}$,
\begin{eqnarray}
 \label{23VII.1}
 &
 \partial_r\gamma_{AB} - 2 r s_{AB} =
    \sum_{i=4}^N i r^{i-1} \stackrel{(i)}{s}_{\!\!AB}
        +O_{\ell-1}(r^{\lambda-1})
  \;,
  &
\\
 \label{23VII.2}
 &
\frac{1}{2} \gamma^{AB}\partial_r \gamma_{AB}
- \frac {n-1}r = \sum_{i=1}^{N-3} r^{i}
 \stackrel{(i)}\psi +O_{\ell-1}(r^{\lambda-3})
  \;,
  &
\end{eqnarray}
where $\gamma^{AB}$ is the matrix inverse of
$\gamma_{AB}$. Further, for some smooth tensors
$\stackrel{(i)}{\sigma}_{\! AB}$ on $S^{n-1}$,
\begin{equation}
 \label{27IIII10.1}
 \gamma_{AC}\sigma^C{}_B := \frac{1}{2}\partial_r\gamma_{AB}
 -\frac{1}{2}\frac{ \gamma^{CD}\partial_r \gamma_{CD}}{n-1}\, \gamma_{AB}
 =  \sum_{i=3}^{N-1} r^{i} \stackrel{(i)}{\sigma}_{\!AB}
       +O_{\ell-1}(r^{\lambda-1})
  \;;
\end{equation}
note that $\stackrel{(3)}\sigma_{AB}$ is the
$s_{AB}$--trace-free part of ${\stackrel{(4)}s}{}_{\!AB}$. This
leads to, for some functions $\stackrel{(i)}f$ on $S^{n-1}$,
\begin{equation}
 \label{30VI.3}
 |\sigma|^2 := \sigma^A{}_B \, \sigma^B{}_A=
  \sum_{i=2}^{N-2} r^{i} \stackrel{(i)}f+O_{\ell-1}(r^{\lambda-2})
  \;,
\end{equation}
where $\sigma^A{}_B$ has been defined in the left-hand
side of \eqref{27IIII10.1}. Let
\begin{equation*}
y:=\frac{n-1}{\tau}
 \;,
\end{equation*}
where $\tau$ is the divergence of $C_O$ (sometimes denoted by
$\theta$ in the literature; cf., e.g.,
\cite{galloway-nullsplitting}):
\begin{equation*}
 \tau:= \frac 12 \overline{g^{AB} \partial_r g_{AB}}
 \;.
\end{equation*}
In terms of $y$, the vacuum Raychadhuri equation
$\overline{R_{11}}\equiv \overline{R_{\mu\nu}} \ell^\mu
\ell^\nu =0$, where $\ell^\nu$ is a null tangent to the
generators of $C_O$, reads
\begin{equation}
 \label{3V.1}
y^{\prime}=1+\frac{1}{n-1}|\sigma|^{2}y^{2}
 \;.
\end{equation}
Using known arguments (compare~\cite{AndChDiss,ChLengardnwe}
and~\cite[Lemma~8.2]{FriedrichCMP86}), there exist functions
$\stackrel{(i)}y\in C^\infty(S^{n-1})$ such that
\begin{equation*}
 y = r + \sum_{i=5}^{N+1}r^i\stackrel{(i)}y+ O_{\ell-1}(r^{\lambda +1})
 \;.
\end{equation*}
We rewrite this as
\begin{equation*}
 y = r(1+\delta y)\;,
\ \text{with} \
\delta y = \sum_{i=4}^{N}r^i\stackrel{(i+1)}y+ O_{\ell-1}(r^{\lambda})
         = O(r^4)
 \;,
\end{equation*}
so that
\begin{equation*}
 \tau = \frac{n-1}y = \frac{n-1}{r(1+\delta y)}=
        \frac{n-1}{r}\left( 1 - \frac{\delta y} {1+\delta y} \right)
 \;.
\end{equation*}
Using this last formula, it is easy to establish existence of
functions $\stackrel{(i)}\tau \in C^\infty(S^{n-1})$ such that
\begin{equation}
 \label{30VI.4}
 \tau - \frac{n-1}r
 = \sum_{i=3}^{N-1}r^i\stackrel{(i)}\tau+ O_{\ell-1}(r^{\lambda -1})
 \;.
\end{equation}
Let us write
\begin{equation}
 \label{8V.2}
\overline{g}_{AB}=e^{\omega} \gamma _{AB}\; ,
\end{equation}
From
\begin{equation}
 \label{5VI.5}
\tau \equiv
\partial_{1}\log \sqrt{\det \gamma}
+\frac{n-1}{2} \partial_{1}\omega\; ,
\end{equation}
there exist functions $\stackrel{(i)}\omega \in C^\infty(S^{n-1})$
such that:
\begin{equation}
 \label{30VI.8}
 \omega  = \sum_{i=4}^{N}r^i\stackrel{(i)}\omega+ O_{\ell-1}(r^{\lambda} )
 \;.
\end{equation}
Equation (\ref{30VI.8}) implies that there exist smooth tensor
fields $\stackrel{(i)}{g}_{\!\!AB}$ on $S^{n-1}$ such that
\begin{equation}
 \label{30VI.7}
 \overline{g}_{AB}-r^2 s_{AB} =
     \sum_{i=4}^N r^{i} \stackrel{(i)}{g}_{\!\!AB}+O_{\ell-1}(r^\lambda)
 \;.
\end{equation}
Subsequently,
\begin{equation}
 \label{30VI.7a}
 \overline{g}^{AB}-r^{-2} s^{AB} =
     \sum_{i=0}^{N-4} r^{i} \stackrel{(i)}{h}{}^{\!\!AB}+
          O_{\ell-1}(r^{\lambda-4})
 \;,
\end{equation}
for some tensor fields fields
$\stackrel{(i)}{h}{}^{\!\!AB} \in C^\infty(S^{n-1})$.

It is a consequence of the affine-parameterization condition
and of the wave-gauge conditions that the function $\nu_0$
solves the equation (see~\cite{CCM2} for details)
\begin{equation}
 \label{5VI.13}
\partial_1\nu^{0}=
   -\frac{1}{2}\tau\nu^{0}+\frac{1}{2}\overline{g}^{AB}rs_{AB}
 \;.
\end{equation}
Set
\begin{equation}
 \label{30VI.11}
Y:=1-\nu^{0}
 \;,
\end{equation}
then (\ref{5VI.13}) integrates to
\begin{eqnarray}
 Y(r)
  &=&
  r^{-\frac{n-1}{2}}\exp\Big(\frac{1}{2}\int_{0}^{r}\psi(\rho)d\rho\Big)
  \int_{0}^{r}\rho^{\frac{n-1}{2}}\exp\Big(-\frac{1}{2}\int_{0}^{\rho}\psi
(\chi)d\chi\Big)F(\rho)d\rho
 \;,
 \nonumber
\\
 &&
 \label{30VI.12}
\end{eqnarray}
where $\psi:= -\tau + (n-1)/r$ and
\begin{equation}
 \label{23VII.1x}
 F :=\frac{1}{2}(\tau-\overline{g}^{AB}rs_{AB})\equiv
     \frac{1}{2}\{(r^{-2}s^{AB} -\overline{g}^{AB})rs_{AB}-\psi\} = O(r)
 \;,
\end{equation}
We find successively
\begin{eqnarray*}
  \int_0^r \psi(s) ds  &= &
     -\sum_{i=3}^{N-1}\frac{r^{i+1}}{i+1}\stackrel{(i)}\tau
         + O_{\ell-1}(r^{\lambda})
 \;,
\\
 F & = & \sum_{i=1}^{N-3} r^{i}\stackrel{(i)} F
         + O_{\ell-1}(r^{\lambda-3})
 \;,
\end{eqnarray*}
for some $\stackrel{(i)}{F} \in C^\infty(S^{n-1})$. Closer
inspection shows that $\stackrel{(1)} F =\stackrel{(2)} F =0$.
Integrating (\ref{30VI.12}), we conclude that there exist
functions $\stackrel{(i)}{\nu}_0\in C^\infty(S^{n-1})$ such
that
\begin{equation}
 \label{30VI.9}
\nu_0 -1= \sum_{i=4}^{N-2} r^{i} \stackrel{(i)}{\nu}_0
              +O_{\ell-1}(r^{\lambda-2})
 \;.
\end{equation}
If $N<6$ the sum from  four to $N-2$ is understood as zero. One
has a similar formula for $\nu^0-1$.

We pass now to a vector $\xi _A$, defined as
\begin{equation}
 \label{xiAbis}
\xi_{A} :=
 -2\nu^{0}\partial_{1}\nu_{A}+4\nu^{0}\nu_{C}\chi_{A}{}^{C}
+\left(\overline{W}^{0}-\frac{2}{r}\nu^{0}\right)\nu_{A}+
\gamma_{AB}\gamma^{CD}(S_{CD}^{B}-\tilde{\Gamma}_{CD}^{B})
 \;.
\end{equation}
Here the $S^A_{BC}$'s are the Christoffel symbols of the
canonical metric $s_{AB}$ on $S^{n-1}$, $W^\alpha$ is as in
(\ref{WaveGauge0}),  the $\tilde \Gamma^A_{BC}$'s  are the
Christoffel symbols of the $(n-1)$-dimensional metric
$\tilde{g}_{AB}$, with associated covariant derivative operator
$\tilde{\nabla}$, and $\nu^0=1/\nu_0$, while
\begin{equation*}
 \chi^A{}_B := \frac 12 \overline{ g^{AC}\partial_r g_{CB}}
  \;.
\end{equation*}
In terms of $\xi_A$ the equation $R_{1A}=0$ in wave-map
gauge reads
\begin{equation}
 \label{CAfinal2}
 -\frac{1}{2}(\partial_{1}\xi_{A}+\tau\xi_{A})
+ \tilde{\nabla}_{B}\chi_{A}{}^{B}
-\frac{1}{2}\partial_{A}(\tau-\overline{W}_{1}+2\nu^{0}\partial_{1}\nu_{0})
 =0
\; ,
\end{equation}
where  $\bar{W}_1 = \nu_0 \bar{W}^0$. From what has been
proved so far this last equation can be written as
\begin{equation}
  \partial_{1}\xi_{A}+\tau\xi_{A}  =
    \sum_{i=1}^{N-3} r^{i}\stackrel{(i)} f_{\!A}+ O_{\ell-2}(r^{\lambda-3})
 \;,
\label{CAfinal2x}
\end{equation}
for some $\stackrel{(i)}{f_{\!A}}\in C^\infty(S^{n-1})$. The
unique solution of this equation with the relevant behaviour at
the origin satisfies
\begin{equation}
   \xi_{A} = \sum_{i=2}^{N-2} r^{i}\stackrel{(i)} \xi_{\!A}
                + O_{\ell-2}(r^{\lambda-2})
 \;,
\label{CAfinal2y}
\end{equation}
for some $\stackrel{(i)}{\xi_{\!A}}\in C^\infty(S^{n-1})$.
Viewing (\ref{xiAbis}) as an equation for $\nu_A$, we obtain
\begin{equation}
   \nu_{A} =  \sum_{i=3}^{N-1} r^{i}\stackrel{(i)} \nu_{\!A}
                 + O_{\ell-2}(r^{\lambda-1})
 \;,
\label{CAfinal2z}
\end{equation}
for some $\stackrel{(i)}{\nu_{\!A}}\in C^\infty(S^{n-1})$.

We finally need to integrate the third wave-map-gauge
constraint, $G_{10}=0$, where $G_{\mu\nu}$ is the Einstein
tensor:
\begin{eqnarray}
 \label{19VIII.1}
 &&  2\partial_{1}^{2}\overline{g}^{11}
    +3\tau \partial_{1}\overline{g}^{11}
    + (\partial_{1}\tau +\tau^{2})\overline{g}^{11}
     \\
\notag
&& \quad
+2(\partial_{1}+\tau)\overline{W}^{1}
+\tilde{R}-\frac{1}{2}\overline{g}^{AB}\xi_{A}\xi_{B}
+\overline{g}^{AB}\tilde{\nabla}_{A}\xi_{B}
=  0 .
\end{eqnarray}
We need the formula, for some functions
$\stackrel{(i)}s \in C^\infty(S^{n-1})$,
\begin{eqnarray}
 \label{2IX.1x}
\tilde{R} &=& \frac{(n-1)(n-2)}{r^2}
 + \sum_{i=0}^{N-4}r^i\stackrel{(i)}s + O_{\ell-3}(r^{\lambda-4})
 \;,
\end{eqnarray}
and the result, for a Minkowski target,
\begin{eqnarray}
 \overline{W}^1 &=& \overline{W}^0 = -\frac{n-1}{r}+ \sum_{i=3}^{N-3}r^i\stackrel{(i)}w + O_{\ell-1}(r^{\lambda-3})
  \label{2IX.2a}
\\
 &=&   -\frac{n-1}{r}+O(r^3)
 \;,
\nonumber
\end{eqnarray}
with $\stackrel{(i)}w \in C^\infty(S^{n-1})$. Hence
\begin{eqnarray}
2(\partial_1+\tau)\overline{W}^1
 &=&
 -2\frac{(n-1)(n-2)}{r^2}+ \sum_{i=2}^{N-4}r^i\stackrel{(i)}v + O_{\ell-1}(r^{\lambda-4})
  \label{2IX.3a}
  \\
 &=&
 -2\frac{(n-1)(n-2)}{r^2}+O(r^2)
 \;,
\nonumber
\end{eqnarray}
for some functions $\stackrel{(i)}v \in C^\infty(S^{n-1})$. A
straightforward analysis of the remaining terms in
(\ref{19VIII.1})  shows that solutions with the required
asymptotics satisfy, for some functions $\stackrel{(i)}g \in
C^\infty(S^{n-1})$,
\begin{eqnarray}\overline{g}^{11}
 &=&
 1+ \sum_{i=2}^{N-2}r^i\stackrel{(i)}g + O_{\ell-3}(r^{\lambda-2})
  \;.
  \label{2IX.4x}
\end{eqnarray}

In order to apply Dossa's existence theory from~\cite{DossaAHP}
to the wave-map gauge reduced Einstein equations, we transform
the metric to coordinates $(y^\mu)$, regular near the tip of
the light-cone, defined as
\begin{equation*}
 y^0 = x^1 - x^0\;, \quad y^i = x^1 \Theta^i(x^A)
 \;,
\end{equation*}
where $\Theta^i(x^A)\in S^{n-1}\subset {\mathbf{R}}^n$. In
these coordinates our wave-map-reduced Einstein equations
become the usual harmonically-reduced equations, which have the
right structure for the existence results
in~\cite{DossaAHP}. One finds that there exist functions
$\stackrel{(i)}g_{\!\mu\nu} \in C^\infty(S^{n-1})$ such that
\begin{equation*}
 \underline{\overline{g}_{\mu\nu}} - \underline{\eta_{\mu \nu}}
 = \sum_{i=2}^{N-2} r^i{}\underline{\stackrel{(i)}g_{\!\mu\nu}}
       + O_{\ell-3}(r^{\lambda-2})
 \;.
\end{equation*}
This proves the first part of Theorem~\ref{T23VII.1}.

Now, for
general $\stackrel{(i)}s_{\!AB}$'s the sum $\sum_{i=2}^{N-2}
r^i\underline{\stackrel{(i)}g_{\!\mu\nu}}$ will \emph{not} be
the restriction of a polynomial to the light-cone. Assume,
however, that this is the case. Then the remainder term in the
difference $\underline{\overline{g}_{\mu\nu}} - \underline{\eta_{\mu
\nu}}$ will be of the differentiability class $F^m(C^T_O)$ with
$m>n/2+1$, as required by Dossa for existence~\cite{DossaAHP},
provided that
\begin{equation}
\label{19VII.1}
  \lambda \ge 2m +1 > n+3
 \;.
\end{equation}
The solution $\underline{g_{\mu \nu}} - \underline{\eta_{\mu
\nu}}$ is then in Dossa's space $\tilde F ^m(Y_O^T)$ for some
$T>0$, which embeds into $W^{m -n/2,\infty}(C_O^T)\subset
W^{1,\infty}(C_O^T)$ leading to, for small $|t|+r$,
\begin{equation}
 \label{10VI.1x}
g_{\mu \nu} - \eta_{\mu \nu} = O_1(|t|+r)
 \;, \quad
 \partial_\sigma g_{\mu\nu} =O(1)
 \;.
\end{equation}
In the coordinates $(y^\mu)$ the harmonicity vector can be
calculated using the usual formula,
\begin{equation*}
 \overline{H}^\alpha =
 \frac {1}{\sqrt {|\det g_{\mu \nu}|}}
        \partial_\beta( \sqrt {|\det g_{\mu \nu}|} g^{\alpha \beta})
 \;,
\end{equation*}
which has bounded components in the $(y^\mu)$ coordinate system
near the tip of the cone. The Bianchi identity (which, in its
simplest version, requires a $C^3$ metric; this raises the
differentiability threshold to $m>n/2+3$) together with the
arguments in~\cite{CCM2} show then that
\begin{equation*}
 \overline{H}^\mu = 0
 \;,
\end{equation*}
and thus the solution of the wave-map reduced Einstein
equations, obtained from Dossa's theorem, is also a solution of
the vacuum Einstein equations.
\hfill $\Box$

\bigskip
\noindent{\textsc{Acknowledgements:}  {PTC and YCB are grateful
to the Mittag-Leffler Institute, Djursholm, Sweden, for
hospitality and financial support during part of work on this
paper. They acknowledge useful discussions with Vincent
Moncrief, as well as comments from Roger Tagn\'{e} Wafo. YCB
wishes to thank Thibault Damour for making available his
detailed manuscript calculations   in the case $n=3$ leading to
equations (22) of~\cite{DamourSchmidt}. JMM thanks OxPDE for
hospitality. }}

\def\polhk#1{\setbox0=\hbox{#1}{\ooalign{\hidewidth
  \lower1.5ex\hbox{`}\hidewidth\crcr\unhbox0}}}
  \def\polhk#1{\setbox0=\hbox{#1}{\ooalign{\hidewidth
  \lower1.5ex\hbox{`}\hidewidth\crcr\unhbox0}}} \def\cprime{$'$}
  \def\cprime{$'$} \def\cprime{$'$} \def\cprime{$'$}
\providecommand{\bysame}{\leavevmode\hbox
to3em{\hrulefill}\thinspace}
\providecommand{\MR}{\relax\ifhmode\unskip\space\fi MR }
\providecommand{\MRhref}[2]{%
  \href{http://www.ams.org/mathscinet-getitem?mr=#1}{#2}
} \providecommand{\href}[2]{#2}

\end{document}